\documentclass[%
onecolumn,
superscriptaddress,
preprintnumbers,
nofootinbib,
nobibnotes,
floatfix,
amsmath,
amssymb,
amsfonts,
aps,
notitlepage
]{revtex4-2}

\usepackage[margin=1.25in]{geometry}
\usepackage{amsmath,amssymb,amsfonts}
\usepackage{amsthm}
\usepackage{graphicx}
\usepackage{color}
\usepackage{mathtools}

\usepackage{siunitx}
\usepackage[bookmarksnumbered,raiselinks,breaklinks]{hyperref}
\hypersetup{
    colorlinks,%
    citecolor=blue,%
    filecolor=blue,%
    linkcolor=blue,%
    urlcolor=blue
}

\graphicspath{{./}}

\newtheorem{definition}{Definition}[section]
\newtheorem{proposition}{Proposition}[section]

\newcommand*{\real}{\ensuremath{ \mathbb{R} }}
\newcommand*{\del}{\nabla}
\newcommand*{\defeq}{\coloneqq}
\newcommand*{\from}{\colon}
\newcommand*{\where}{\mid}
\newcommand{\restr}[1]{|_{#1}}
\newcommand\dd{\mathrm{d}} 
\newcommand*\dif{\mathop{}\!\mathrm{d}} 
\newcommand*\bb[1]{\mathbf{#1}}
\newcommand*\vv[1]{\boldsymbol{#1}}
\DeclarePairedDelimiter\abs{\lvert}{\rvert}
\DeclarePairedDelimiter\norm{\lVert}{\rVert}

\makeatletter
\renewcommand\thesection{\@arabic\c@section}
\renewcommand\thesubsection{(\@alph\c@subsection)}
\renewcommand\p@subsection{\thesection}
\def\@seccntformat#1{\@ifundefined{#1@cntformat}%
   {\csname the#1\endcsname\space}
   {\csname #1@cntformat\endcsname}}
\newcommand\section@cntformat{\thesection.\space}       
\newcommand\subsection@cntformat{\thesubsection\space} 
\makeatother

\begin{document}

\title{Localization in musical steelpans}

\author{Petur Bryde}
\affiliation{Paulson School of Engineering and Applied Sciences, Harvard University, Cambridge, MA 02138.}
\author{L.\ Mahadevan}
\email{lmahadev@g.harvard.edu}
\affiliation{Paulson School of Engineering and Applied Sciences, Harvard University, Cambridge, MA 02138.}
\affiliation{Department of Physics, Harvard University, Cambridge, MA 02138.}
\affiliation{Department of Organismic and Evolutionary Biology, Harvard University, Cambridge, MA 02138.}

\begin{abstract}
The steelpan is a pitched percussion instrument that takes the form of a concave bowl with
several localized dimpled regions of varying curvature. Each of these localized zones,
called notes, can vibrate independently when struck, and produces a sustained tone of a
well-defined pitch. While the association of the localized zones with individual notes has
long been known and exploited, the relationship between the shell geometry and the strength
of the mode confinement remains unclear. Here, we explore the spectral properties of the
steelpan modeled as a vibrating elastic shell. To characterize the resulting eigenvalue
problem, we generalize a recently developed theory of localization landscapes for scalar
elliptic operators to the vector-valued case, and predict the location of confined
eigenmodes by solving a Poisson problem. A finite element discretization  of the shell shows
that the localization strength is determined by the difference in curvature between the note
and the surrounding bowl. In addition to providing an explanation for how a steelpan
operates as a two-dimensional xylophone, our study provides a geometric principle for
designing localized modes in elastic shells.
\end{abstract}

\maketitle

\section{Introduction}

\begin{figure}
    \centering
    \includegraphics[width=0.8\textwidth, trim={0cm 0.00cm 0.00cm 0.00cm}, clip]{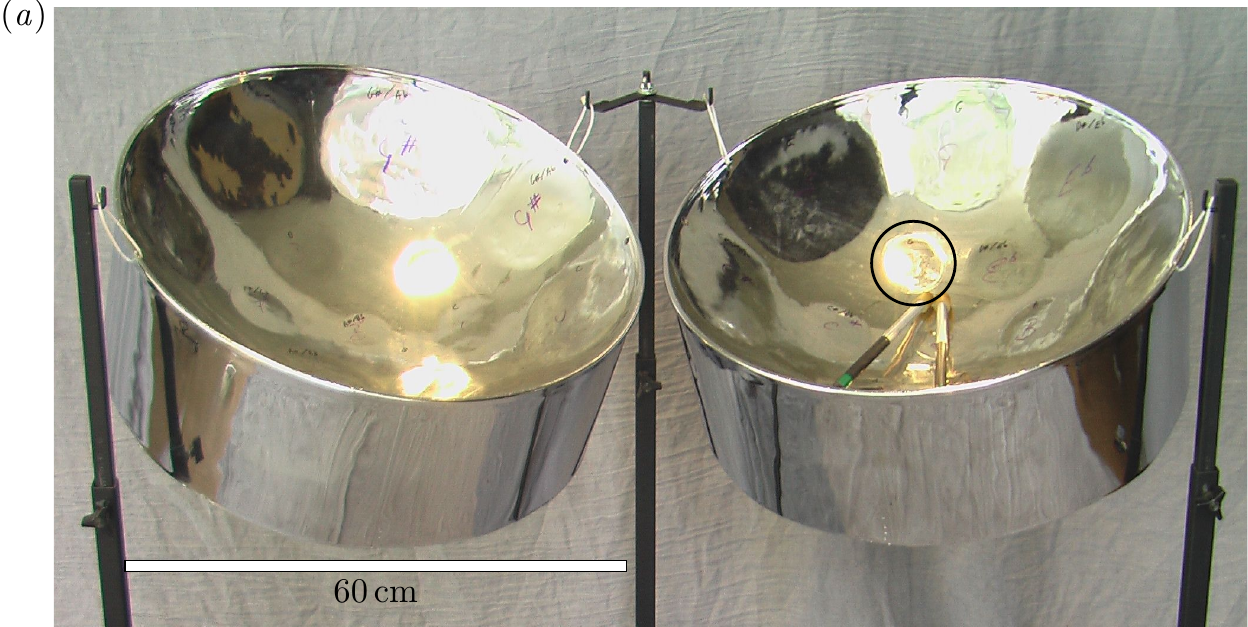}
    \includegraphics[width=0.8\textwidth, trim={0cm 0.00cm 0.00cm 0.00cm}, clip]{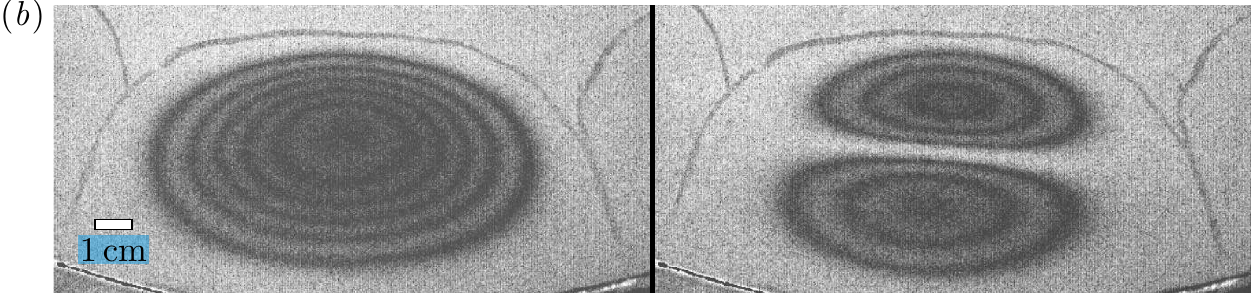}
    \caption{
        (\textit{a}) A ``double second'' steelpan with two types of note regions:
        The outer notes in the shape of rounded rectangles and elliptical inner notes. One
        of the inner notes is indicated by a black circle. Adapted from
        Vetter~\cite{panphoto},
        \copyright 2003 Roger Vetter, courtesy of the Grinnell College Musical Instrument
        Collection.
        (\textit{b}) The first (left) and second (right) localized eigenmodes of a C\(_4\) note on a
        ``tenor'' steelpan imaged using time-averaged TV holography.
        Adapted from Morrison~\cite{morrisonthesis} with permission from A.C.\ Morrison.
    }
\label{panimages}
\end{figure}

When an elastic structure such as a beam, plate or shell of uniform curvature is struck, the
resulting vibration quickly propagates as a wave through the entire system. In contrast, a
flat portion of a shell surrounded by a region of higher curvature may support localized
vibrational modes, i.e.\ stationary waves which are confined to a small subregion. This is
the basic principle behind the steelpan, a pitched percussion instrument originating from
Caribbean approximately a century ago \cite{rossing1996}. It consists of a concave playing
surface (referred to as the bowl) joined at the boundary to a cylindrical ``skirt'', as
shown in figure~\ref{panimages}\textit{a}. The playing surface has a number of flat or
slightly concave regions which are able to vibrate independently, each with its own pitch
(frequency) determined by its size and shape. These regions are called the notes of the pan,
as each one is designed to resonate at a frequency corresponding to a certain note on the
musical scale, thus making the steelpan a two-dimensional xylophone.

Steelpans were originally made from standard 55-gallon steel drums. The bottom face of the
drum is hammered into a concave bowl shape and the notes are defined by locally raising and
flattening the bowl, creating regions of lower curvature. A steelpan may have anywhere
between three to upwards of 30 notes depending on the desired range. In most pans, the notes
are arranged in concentric circles, with the inner notes either circular or elliptical while
the outer notes resemble rounded rectangles (figure~\ref{panimages}\textit{a}). Part or all
of the cylindrical side is retained as the skirt, which acts as an acoustic baffle, i.e.\ it
prevents the cancellation of sound from the two sides of the bowl~\cite{rossing1996}. The
steelpan and the crafting process have received considerable interest across multiple
fields, including acoustics, mechanics and material science ~\cite{kronman1992}. Studies of
the steelpan have predominantly been experimental, and focus on either the materials science
and metallurgy of the construction processes~\cite{murr1999}, or the measurements of spectra
and mode shapes~\cite{rossing1996,rossingexp1,maloneyconfinement,morrisonthesis} and of the
sound field~\cite{copeland2005}.

The pan is played by striking the notes with a soft-tipped mallet. However, the mechanism
underlying the mode confinement in steelpan notes is still incomplete. Steelpan makers
conventionally outline each note with chisel marks and mode confinement has sometimes been
ascribed to this grooving process~\cite{kronman1992, murr1999}. This explanation was
challenged by Maloney, Barlow, and Woodhouse, who proposed that mode confinement is a
consequence of the steelpan geometry~\cite{maloneyconfinement}. In a numerical and
experimental modal analysis, they found that a flat circular note region on a hemispherical
pan supports localized modes, suggesting that confinement is influenced by bowl
curvature~\cite{maloneyconfinement}. Later experimental studies using holographic
interferometry reveal a number of normal modes that are completely localized to a single
note region~\cite{morrisonthesis}; the mode shapes are qualitatively similar to those of a
flat plate with the same dimensions. In general, the modes are designated by a pair of
numbers \((m ,n)\), where \(m\) is the number of radial nodal lines and \(n\) is the number
of nodal lines perpendicular to the radial direction, e.g. the frequency of the fundamental,
or \((0,0)\) mode determines the pitch, and makers of the steelpan carefully control the
geometry to achieve two or three harmonic overtones. To visualize this,
Figure~\ref{panimages}\textit{b} shows experimental images of the \((0,0)\) and \((0,1)\)
modes of a C\(_4\) note, both displaying strong localization.

A similar mode confinement effect is observed in a related instrument, the musical saw,
which consists of a strip of metal such as the blade of a regular handsaw. When bent into an
`S-shape', a number of localized vibrational modes emerge at the inflection line, where the
non-vanishing principal curvature changes sign~\cite{ScottWoodhouse}.
Recent work~\cite{shankar2022}, quantified how the localization strength varies with the
curvature gradient. Moreover, our analysis suggests that these classical confined modes are
topologically protected, analogously to boundary modes in quantum-mechanical topological
insulators.

Here, we characterize how localization in elastic shells arises from variations in the
curvature of doubly curved shells, and characterize the strength of mode localization in the
steelpan as a function of bowl curvatures. We model the steelpan as a thin linear elastic
shell with an inhomogeneous curvature, whose dynamics is governed by a set of coupled
partial differential equations. To understand the vibrational properties of the system, we
have to solve an eigenvalue problem for an idealized version of the pan geometry.

Here, instead we introduce an alternative approach based on a generalization of the
localization landscape method of Filoche and Mayboroda~\cite{filoche2012}. Originally
developed as a geometric alternative to explain Anderson localization \cite{anderson} in
quantum mechanical systems, one computes the landscape function associated with the system,
whose peaks coincide with the possible localization regions. The landscape can be
approximated by solving a single Poisson-like problem for the same operator, a computation
which is considerably less expensive than solving the eigenvalue problem. The approach has
been successfully used to analyze localized modes in various systems described by scalar
equations \cite{filoche2017,arnold2016,arnold2019a,arnold2019b}. Our generalization of the
technique is applicable to elliptic systems of partial differential equations, which
describe a wide variety of physical phenomena, including shell structures. We demonstrate
the method for a steelpan-like shell with realistic note shapes.

The organization of the paper is as follows. In \S\ref{sec:localization}, the landscape
theory for scalar PDE is reviewed and the vector landscape is introduced. In
\S\ref{sec:naghdi} we discuss the shell theory which we use to model the steelpan. In
\S\ref{sec:methodology} describes the methodology of the modal analysis, including the
simplified pan geometries and the numerical methods. The results are presented in
\S\ref{sec:results}.

\section{Vector localization landscape for elliptic systems}
\label{sec:localization}

\subsection{Review of localization landscape theory for scalar equations}
Localization is a phenomenon exhibited by some vibrating systems where a standing wave is
concentrated inside a small part of the domain and almost vanishing outside of it. As a
result, a disturbance at one point of the medium need not propagate to the rest of the
system.
In the context of quantum mechanics, a localized wavefunction describes a particle which
is confined to one region. A familiar example is localization due to a potential well,
but the effect is observed in many less obvious cases.
Localization may stem from the domain geometry, for instance due to a rough or irregular
boundary, or when the system is composed of several weakly connected subdomains~\cite{grebenkov2013}.
Confined modes can also arise in a disordered medium.
The principal example of this is Anderson localization, where electron wavefunctions in a
crystal are localized in the presence of a sufficiently rough or disordered
potential~\cite{anderson}.

To study localization in a general setting, we consider an elliptic partial differential
operator \(L\) on a domain \(\Omega\) in \(\real^n\)
and the associated eigenvalue problem \(L w = \lambda w\).
For instance, if \(L = -\Delta + V\) is a Schrödinger operator where the potential \(V\)
is piecewise constant with randomly chosen values, we obtain a model for Anderson
localization.
In general, it is not obvious whether any modes of \(L\) are localized or what the
localization regions are. The recently developed localization landscape (LL) theory of
Filoche and Mayboroda~\cite{filoche2012} provides a way to predict the confinement
properties of low-frequency modes without solving the full eigenvalue problem.
For a symmetric elliptic operator \(L\) acting on scalar functions, they defined the
\emph{localization landscape} as the function \(\mathcal{L} \from \Omega \to \real\) given by
\(\mathcal{L}(x) = \int_{\Omega} \abs{G(x,y)} \dif y\), where \(G\) is the Green's function
of \(L\) which satisfies the same boundary conditions as imposed in the eigenvalue problem.
In systems which support low-frequency localized modes, the landscape has one or more peaks,
coinciding with the localization regions, separated by valleys, where the eigenfunctions are
necessarily small. This is made precise by the following inequality, satisfied by each
eigenpair
\((\lambda, w)\):
\begin{equation}
    \label{lscapeineq}
    \frac{ \abs{w(x)} }{\norm{w}_{L^{\infty}(\Omega)}} \leq \lambda \mathcal{L}(x),
    \qquad \forall x \in \Omega.
\end{equation}
The proof is remarkably short in the case when \(w\) is smooth (see~\cite{filoche2012}):
from the definition of the Green's function and the symmetry of \(L\), we have
\begin{equation}
    w(x) = \int_{\Omega} L_y G(x,y) w(y) \dif y =
    \int_{\Omega} G(x,y) L_y w(y) \dif y = \int_{\Omega} G(x,y) \lambda w(y) \dif y,
\end{equation}
and thus
\begin{equation}
    \abs{w(x)} \leq \lambda \norm{w}_{L^{\infty}(\Omega)} \int_{\Omega} \abs{G(x,y)} \dif y.
\end{equation}
If \(w\) is not necessarily smooth, we replace \(w(x)\) by
\(\int_{\Omega} w(z) \varphi_{\varepsilon}(x-z) \dif z\) where \(\varphi_{\varepsilon}\) is
a mollifier.

The inequality~\eqref{lscapeineq} says, roughly, that \(w\) is concentrated in the superlevel set
\(E(\lambda) = \{x \in \Omega \where \mathcal{L}(x) \leq \lambda \}\).
Note that \(E(\lambda)\) may have several connected components, each encompassing a
peak of the landscape \(\mathcal{L}\).
Therefore, the inequality~\eqref{lscapeineq} does not by itself guarantee that \(w\) is
localized; instead it may be a linear combination of localized functions, each supported
in a single connected component of \(E(\lambda)\).
However, numerical experiments~\cite{filoche2012,arnold2016} have shown that
typically, each eigenfunction is localized near a single peak of \(\mathcal{L}\),
with the following exceptions:
(i) two eigenfunctions with nearly the same eigenvalue can share a peak, and
(ii) an eigenfunction may be spread over two or more peaks that are sufficiently close
together. This much stronger result has been made precise and shown rigorously for
elliptic operators of the form
\(L = -{\operatorname{div}} A(x) \nabla + V(x)\)~\cite[Theorem 5.1]{arnold2019b}.

It should be noted that when the Green's function is non-negative, as is the case when
\(L\) is of second order, \(\mathcal{L}\) is precisely the solution \(u\) to the boundary
value problem \(L u = 1\) on \(\Omega\), which simplifies its computation dramatically.
For higher order operators, this is generally not the case; indeed, this condition fails
even for the bilaplacian on certain domains~\cite{grunau2014}.
However, if the negative part of \(G\) is relatively small, one can approximate the
landscape by the solution to \(L u = 1\) and still obtain qualitatively correct results.

The localization landscape theory has found applications in semiconductor
physics~\cite{filoche2017} and biochemistry~\cite{chalopin2019} among other fields. However,
many physical systems are described by systems of PDE, to which the existing landscape
theory is not applicable. In particular, this includes the equations of linear elasticity
and the shell equations considered in this paper. In the following section, we state an
appropriate extension of the LL theory to elliptic systems by defining a vector-valued
landscape \(\vv{\mathcal{L}}\) and proving a generalization of the
inequality~\eqref{lscapeineq}.

\subsection{Generalization to elliptic systems}
To simplify the exposition, we assume homogeneous Dirichlet boundary conditions throughout.
In the following, \(\norm{\cdot}\) and \(\langle \cdot, \cdot \rangle\) denote
respectively the Euclidean norm and inner product of vectors or matrices considered
as elements of \(\real^{m \times n}\).

Let \(\Omega \subset \real^{n}\) with \(n \geq 2\) be a bounded domain and let \(L\) be
a second order, symmetric elliptic operator which acts on vector-valued functions
\(\bb{u} \from \Omega \to \real^{m}\) according to
\begin{equation}
    \label{strongoperator}
    (L \bb{u})_i = -\partial_{\alpha} \left(A^{\alpha \beta}_{ij}
    \partial_{\beta} u^{j}\right) + B_{ij} u^j,
\end{equation}
where \(A^{\alpha \beta}_{ij}\), \(B_{ij}\) are bounded, measurable functions.
Since localization often arises in rough domains or for systems with highly irregular
coefficients, we impose no further regularity restrictions on \(\Omega\) or the
coefficients.
The operator should be understood in the weak sense, as follows: Let \(H^1(\Omega)\)
denote the usual Sobolev space of functions \(f \in L^2(\Omega)\) with weak derivatives
in \(L^2(\Omega)\). The space \(H^1_0(\Omega)\) is the completion in \(H^1(\Omega)\) of
\(C_{c}^{\infty}(\Omega)\), the set of smooth functions with compact support. Functions in
\(H^1_0(\Omega)\) are said to satisfy Dirichlet boundary condition in the weak sense.
The action of the operator \(L\) and the corresponding bilinear form \(a\) is given by
\begin{equation}
    \label{bilinearform}
    (L \bb{u}, \bb{v}) \defeq a(\bb{u}, \bb{v}) = \int_{\Omega}
    A^{\alpha \beta}_{ij}(x) \partial_{\beta} u^{j} \partial_{\alpha} v^{i} +
    B_{ij} u^{i} v^{j} \dif x
\end{equation}
for any \(\bb{u}\), \(\bb{v} \in \bb{H}^1_0 \defeq (H^1_0(\Omega))^m\).
For a vector field \(\bb{f}\), we say that \(\bb{u}\) is a weak solution of the Dirichlet problem
\(L \bb{u} = \bb{f}\), \(\bb{u}\restr{\partial \Omega} = 0\) if \(\bb{u} \in \bb{H}^1_0\) and
\begin{equation}
    \label{weaksolution}
    a(\bb{u}, \bb{v}) = (\bb{f}, \bb{v}) \quad
    \forall \bb{v} \in \bb{H}^1_0,
\end{equation}
where \((\bb{f}, \bb{v}) = \int_{\Omega} \bb{f} \cdot \bb{v} \dif x\) denotes the inner
product on \((L^2(\Omega))^m\).

The assumption that \(L\) is symmetric, meaning that
\(a(\bb{u}, \bb{v}) = a(\bb{v}, \bb{u})\) for all \(\bb{u}, \bb{v} \in \bb{H}^1_0\),
is equivalent to the condition
\begin{equation}
    \label{symmetrycondition}
    A^{\alpha \beta}_{i j} = A^{\beta \alpha}_{j i}, \quad B_{i j} = B_{j i}.
\end{equation}
In addition, we assume that the bilinear form \(a\) is coercive on \(\bb{H}^1_0\), i.e.\ that
for some \(\mu > 0\),
\begin{equation}
    \label{coercive}
    a(\bb{u}, \bb{u}) \geq \mu \norm{\bb{u}}^2_{\bb{H}^1_0}
    = \mu \int_{\Omega} \big( \norm{\nabla \bb{u}}^2 + \norm{\bb{u}}^2 \big) \dif x,
        \quad \forall \bb{u} \in \bb{H}^1_0.
\end{equation}
This holds, for example, if the coefficients satisfy the strong ellipticity condition
(see e.g.~\cite[chapter~13]{chipot2009}),
\begin{equation}
    \label{ellipticity}
    \begin{aligned}
        &A^{\alpha \beta}_{i j}(x) \xi^{i}_{\alpha} \xi^{j}_{\beta} \geq
        c \norm{\vv{\xi}}^2, &
                      &\forall \vv{\xi} \in \real^{m \times n},\\
        &B_{ij}(x) \zeta^{i} \zeta^{j} \geq 0, &
                      &\forall \vv{\zeta} \in \real^m,
    \end{aligned}
\end{equation}
for some \(c > 0\) (see also~\cite{davey2018} for weaker but more technical conditions).
Under these assumptions, there exists a unique weak solution to the boundary value
problem~\eqref{weaksolution} for any \(\bb{f} \in [L^2(\Omega)]^m\).
As an example, we note that our setting includes in particular the Lamé operator
\(L \bb{u} = - \partial_{\alpha} (H^{\alpha \beta \mu \nu} E(\bb{u})_{\mu \nu}) \)
of three-dimensional linear elasticity (see \S\ref{subsec:eom}),
and the Naghdi shell operator of equation~\eqref{naghdi}.

In continuum mechanics, one often encounters generalized eigenvalue problems of the
form (cf.\ equation~\eqref{naghdieigen})
\begin{equation}
    \label{generalizedeigen}
    a(\bb{u}, \bb{v}) = \lambda m(\bb{u}, \bb{v}), \quad \forall \bb{v} \in \bb{H}^1_0.
\end{equation}
where the bilinear form \(m\) is given by
\(m(\bb{u}, \bb{v}) = \int_{\Omega} \langle \bb{M}(x) \bb{u}(x), \bb{v}(x) \rangle \dif x\).
We assume that the matrix-valued function \(\bb{M}\) is positive semidefinite for all \(x\)
and with coefficients in \(L^{\infty}(\Omega)\).
Our goal is to obtain pointwise bounds for the generalized eigenfunctions in terms of
integrals of the Green matrix of the system. We repeat here the definition of the Green
matrix given in~\cite{davey2018}, specialized to our case. In the following, \(\bb{e}_k\)
denotes the \(k\)th standard unit vector in \(\real^{m}\) and
\(B_r(y) \defeq \{x \in \real^n \where \norm{x-y} < r\}\) is the open ball with center
\(y\) and radius \(r\).
\begin{definition}
    \label{greendef}
    We say that the matrix function \(\bb{G}(x,y) = (G_{ij}(x,y))_{i,j=1}^{m}\)
    defined on \(\{(x,y) \in \Omega \times \Omega \where x \neq y\}\) is the Dirichlet
    Green matrix of \(L\) if the following holds:
    \begin{enumerate}
        \item
            For all \(y \in \Omega\),
            \(\bb{G}(\cdot, y)\) is locally integrable and satisfies
            \(L \bb{G}(\cdot, y) = \delta_y I\) in the weak sense, i.e.\
            \begin{equation}
                \label{greendelta}
                a(\bb{G}(\cdot, y) \bb{e}_k, \vv{\varphi}) = \varphi^k(y),
                \quad \forall \vv{\varphi} \in C^{\infty}_{c}(\Omega)^m.
            \end{equation}
        \item
            For any \(y \in \Omega\) and \(r > 0\),
            \(\bb{G}(\cdot, y) \in H^{1}(\Omega \setminus B_r(y))^{m \times m}\),
            and \(\bb{G}(\cdot, y)\) vanishes on \(\partial \Omega\) in the sense that
            for every \(\zeta \in C^{\infty}_{c}(\Omega)\) satisfying \(\zeta
            \restr{B_r(y)} \equiv 1\) for some \(r > 0\), we have
            \begin{equation}
                (1-\zeta) \bb{G}(\cdot, y) \in H^{1}_{0}(\Omega \setminus
                B_r(y))^{m \times m}
            \end{equation}
        \item
            For any \(\bb{f} = (f^1, \ldots, f^m)^{T} \in L^{\infty}(\Omega)^m\), the
            function \(\bb{u} = (u^1, \ldots, u^m)^{T}\) given by
            \begin{equation*}
                \bb{u}(x) = \int_{\Omega} \bb{G}(x, y) \bb{f}(y) \dif y
            \end{equation*}
            is the unique weak solution in \(\bb{H}^1_0\) to \(L \bb{u} = \bb{f}\).
    \end{enumerate}
\end{definition}
We note that since \(L\) is assumed to be symmetric, \(\bb{G}\) satisfies the symmetry
relation \(\bb{G}(x, y) = \bb{G}(y, x)^{T}\).
Assume that \((\lambda, \bb{w})\) is an eigenpair, i.e.\ a solution
of~\eqref{generalizedeigen}, and assume moreover that \(\bb{w} \in L^{\infty}(\Omega)^m\).
Then \(\lambda \bb{M} \bb{w} \in L^{\infty}(\Omega)^m\) and from property (iii) of
Definition~\eqref{greendef}, it follows that the function \(\bb{u}\) defined by
\begin{equation*}
    \bb{u}(x) = \int_{\Omega} \bb{G}(x, y) \lambda \bb{M}(y) \bb{w}(y) \dif y
\end{equation*}
is in \(\bb{H}^1_0\) and satisfies \(L \bb{u} = \lambda \bb{M} \bb{w}\) in the sense of
distributions. By uniqueness of solutions, it follows that \(\bb{u} = \bb{w}\).
Interchanging the roles of \(x\) and \(y\) and using the symmetry property of the Green's
matrix, we find for the \(i\)th component of \(\bb{w}\):
\begin{equation*}
    w^i(y) = \lambda \int_{\Omega}
    \langle \bb{G}(y, x) \bb{M}(x) \bb{w}(x), \bb{e}_i \rangle \dif x =
    \lambda \int_{\Omega}
    \langle \bb{M}(x) \bb{w}(x), \bb{G}(x, y) \bb{e}_i \rangle \dif x.
\end{equation*}
Let \(\bb{M}^{\frac{1}{2}}\) denote the unique positive semidefinite square root of \(\bb{M}\).
From the Cauchy--Schwarz inequality in \(\real^m\), we have
\begin{equation*}
    \begin{aligned}
        \abs{w^i(y)} & \leq \lambda \int_{\Omega}
        \abs{ \langle \bb{M}^{\frac{1}{2}}(x) \bb{w}(x), \bb{M}^{\frac{1}{2}} \bb{G}(x, y) \bb{e}_i \rangle } \dif x\\
                     & \leq \lambda \int_{\Omega}
                     \norm{\bb{M}^{\frac{1}{2}}(x) \bb{w}(x)} \norm{\bb{M}^{\frac{1}{2}}(x) \bb{G}(x, y) \bb{e}_i} \dif x\\
                     &\leq
                     \lambda \big\lVert
                         \bb{M}^{\frac{1}{2}} \bb{w}
                     \big\rVert_{L^{\infty}(\Omega)}
                     \int_{\Omega} \norm{\bb{M}^{\frac{1}{2}}(x) \bb{G}(x, y) \bb{e}_i} \dif x.
    \end{aligned}
\end{equation*}
Defining the \emph{vector landscape}
\(\vv{\mathcal{L}} \from \Omega \to \real^m\) as the vector field with components
\begin{equation}
    \label{vectorlandsape}
    \mathcal{L}^i(y) \defeq \int_{\Omega} \norm{\bb{M}^{\frac{1}{2}} \bb{G}(x, y) \bb{e}_i} \dif x.
\end{equation}
we obtain the following generalization of the scalar inequality~\eqref{lscapeineq}:
\begin{proposition}
    \label{prop:landscape}
    Assume a Dirichlet Green's matrix \(\bb{G}\) exists for \(L\) and that \(\bb{w}\) is a
    bounded eigenfunction satisfying~\eqref{generalizedeigen}. Then
    \begin{equation}
        \label{vectorlandscapeineq}
        \Big(\big\lVert \bb{M}^{\frac{1}{2}} \bb{w} \big\rVert_{L^{\infty}(\Omega)}\Big)^{-1} \abs{w^i(x)}
        \leq \lambda \mathcal{L}^i(x), \qquad x \in \Omega.
    \end{equation}
    where \(\vv{\mathcal{L}}\) is defined in equation~\eqref{vectorlandsape}.
\end{proposition}
The interpretation is similar to the scalar case:
The inequality states that each component of \(w^{i}\) is concentrated in the superlevel
set \(E^{i}(\lambda) = \{x \in \Omega \where \mathcal{L}^{i}(x) \leq \lambda \}\)
of the corresponding component of the vector landscape.
Our numerical experiments experiments (\S\ref{results:landscape}) suggest
that, at least for some systems, \(w^{i}\) will in fact be localized near a single peak of
\(\mathcal{L}^{i}\), as in the scalar case. We conjecture that an analogue of theorem 5.1
in~\cite{arnold2019b} holds also for the vector landscape.

As a concrete example, consider the equations of three-dimensional linear elasticity,
equation~\eqref{weakeom}. Then \(\bb{M} = \bb{I}\) is the identity matrix,
and the vector landscape yields a simple upper bound on the norm of the displacement
\(\bb{U}\):
\begin{equation}
    \norm{\bb{U}(x)} \leq \lambda \norm{\vv{\mathcal{L}}(x)}, \qquad
    \mathcal{L}^i(x) = \int_{\Omega} \norm{\bb{G}(x, y) \bb{e}_i} \dif x,
\end{equation}
where the eigenfunction has been normalized so that \(\norm{\bb{U}}\) has
\(L^{\infty}\)-norm 1.
In the remainder of the paper, we will consider the Naghdi eigenvalue problem,
equation~\eqref{naghdieigen}, for which \(\bb{M}\) is the diagonal matrix \(\bb{M} =
\operatorname{diag}(1, 1, 1, 0, 0)\), and we again obtain an upper bound on the norm of
the displacement:
\begin{equation}
    \norm{\bb{u}(x)} \leq \lambda \norm{\vv{\mathcal{L}}(x)}, \qquad
    \mathcal{L}^i(x) = \int_{\Omega} \norm{\tilde{\bb{G}}(x, y) \bb{e}_i} \dif x
\end{equation}
where \(\tilde{\bb{G}}\) is the upper-left \(3 \times 3\) submatrix of \(\bb{G}\) and
\(\bb{e}_i \in \real^3\).
Finally, we note that the statement of Proposition~\ref{prop:landscape} is
easily modified to account for different boundary conditions, namely by replacing the
Dirichlet Green's matrix with one satisfying the same boundary conditions as \(\bb{w}\).

\section{The elastic shell model}
\label{sec:naghdi}

To apply our generalization of the localization landscape to the steelpan modeled
mathematically as a thin elastic shell. The mechanics of three-dimensional structures which
are thin in one direction in comparison with the other two are simplified versions of the 3D
equations of continuum mechanics, and apply to the two-dimensional problem of finding the
deformation of the midsurface. Shell theories can be seen as a generalization of the more
familiar plate theories, which model flat bodies, to structures which may be curved in their
rest state. Like plates, shells are much weaker to bending (i.e.\ isometric deformations)
than they are to stretching and shearing. For shells, however, the membrane (stretching)
strains are generally coupled to the bending and shearing strains due to the curvature of
the midsurface. This has a dramatic effect on the mechanics, allowing shells to resist
applied loads more effectively. The curvature of the midsurface also complicates the
analysis: In contrast to plates, the equations describing in-surface and transverse
displacements are coupled, and one must generally work in curvilinear coordinates. Apart
from exceptional cases, the equations of shell theory can only be solved numerically.

There are a number of different shell theories, each based on different physical
assumptions; see for example \cite{bathechapelle} for an overview. Here we use a specific
model known as the Naghdi shell model~\cite{naghdi1973}, which allows for extensional and
shear deformations of the mid-surface, as well shear and bending deformations transverse to
the mid-surface. These kinematical assumptions are the same as the well-known
Reissner--Mindlin model for thin plates \cite{bathechapelle}, to which the Naghdi model
reduces in the special case of a flat midsurface. The choice of model is motivated by ease
of computation: shearable theories such as the Naghdi model require only \(H^1\)-conforming
finite elements methods, which are relatively simple to implement and are available in many
free software packages. This should be contrasted with models that do not allow for
transverse shear which give rise to weak formulations with solutions in \(H^2\), and thus
call for more complicated \(H^2\)-conforming or discontinuous Galerkin methods.

We start with a derivation of the Naghdi equations for completeness,
following~\cite{bathechapelle}, leading directly to the equations of motion in weak form,
required to apply the finite element method. Since our goal is modal analysis of the
steelpan, we limit the discussion to the linearized theory and clamped boundary conditions.
For simplicity, we consider only the case of an isotropic, homogeneous material of constant
thickness.

\subsection{Differential geometry of a deforming shell}

\label{subsec:geometry}

We adopt the convention that Latin indices denote components of (2D) surface tensors and
take the values $\{1,2\}$, while Greek indices denote components of 3D tensors and take the
values $\{1,2,3\}$; repeated indices are summed as usual. The shell is defined as a
three-dimensional (3D) slender elastic body specified by a reference surface $\mathcal{S}$,
representing the midsurface of the undeformed configuration, and a constant thickness $h$,
which is assumed small compared to the lateral dimensions and radius of curvature of the
shell. The midsurface is parametrized by a map $\bb{X}_0:\Omega_0 \to \real^3$ where the
\emph{reference domain} $\Omega_0$ is a bounded domain in $\real^2$. At any point on the
midsurface, the vectors $\bb{e}_{i}=\partial_i\bb{X}_0$ are linearly independent, and form
the covariant basis for the tangent plane. We also define the unit normal vector
$\bb{e}_3 = (\bb{e}_1\times\bb{e}_2)/\norm{\bb{e}_1\times\bb{e}_2}$.
The undeformed shell body is then parametrized by
\begin{equation}
    \bb{R}_0(\xi^1, \xi^2, \xi^3)=\bb{X}_0(\xi^1, \xi^2) +
    \xi^3\bb{e}_3(\xi^1,\xi^2), \quad (\xi^1, \xi^2,
    \xi^3)\in\Omega \defeq \Omega_0\times\left[-\frac{h}{2}, \frac{h}{2}\right],
\end{equation}
where $\xi^\alpha$ are the local 3D coordinates within the material.
We follow the usual convention of denoting components of tensors with respect to
the covariant basis by superscripts; we call these the \emph{contravariant} components.
Subscript indices indicate \emph{covariant components}, which are components with
respect to the contravariant basis \((\bb{e}^{1}, \bb{e}^{2})\), defined by the
relation \(\bb{e}^{i} \cdot \bb{e}_{j} = \delta^{i}{}_{j}\).

The first fundamental form \(\bb{a}\) and second fundamental form \(\bb{b}\) of the
reference midsurface have the covariant components
\begin{equation}
    a_{ij}=\partial_i\bb{X}_0\cdot\partial_j\bb{X}_0, \quad
    b_{ij}=\bb{e}_3\cdot\partial_i\partial_j\bb{X}_0.
\end{equation}
The reference area element of the surface is $\dd A=\sqrt{\det(\bb{a})}\dd\xi^1\dd\xi^2$.
Indices of midsurface tensors are raised and lowered using \(\bb{a}\). For example,
the components of a vector \(\bb{V} = V^{i} \bb{e}_i = V_i \bb{e}^{i}\) are related by
\(V_i = a_{i j} V^{j}\).
The 3D reference metric tensor
$g_{\alpha\beta}=\partial_\alpha\bb{R}_0\cdot\partial_\beta\bb{R}_0$ has a simple expression
in terms of the fundamental forms:
\begin{equation}
    g_{ij}=a_{ij}-2\xi^3b_{ij}+(\xi^3)^2b_i^kb_{kj}, \quad
    g_{i3}=0, \quad g_{33}=1.\label{3Dmetric}
\end{equation}
Finally, covariant derivatives of midsurface tensors are defined in terms
of the Christoffel symbols \(\Gamma_{ij}^{k} = \partial_{i} \bb{e}_{j} \cdot \bb{e}^{k}\)
of the reference surface, so that e.g.,
\begin{equation}
    \del_i V^j = \partial_i V^j + \Gamma_{i k}^{j} V^k, \qquad
    \del_i V_j = \partial_i V_j - \Gamma_{i j}^{k} V_k.
\end{equation}
for the vector field \(\bb{V}\).

\subsection{Kinematics}
\label{subsec:kinematics}
\begin{figure}
    \centering
    \includegraphics[width=1.0\textwidth, trim={0cm 0.0cm 0.0cm 0.0cm}, clip]{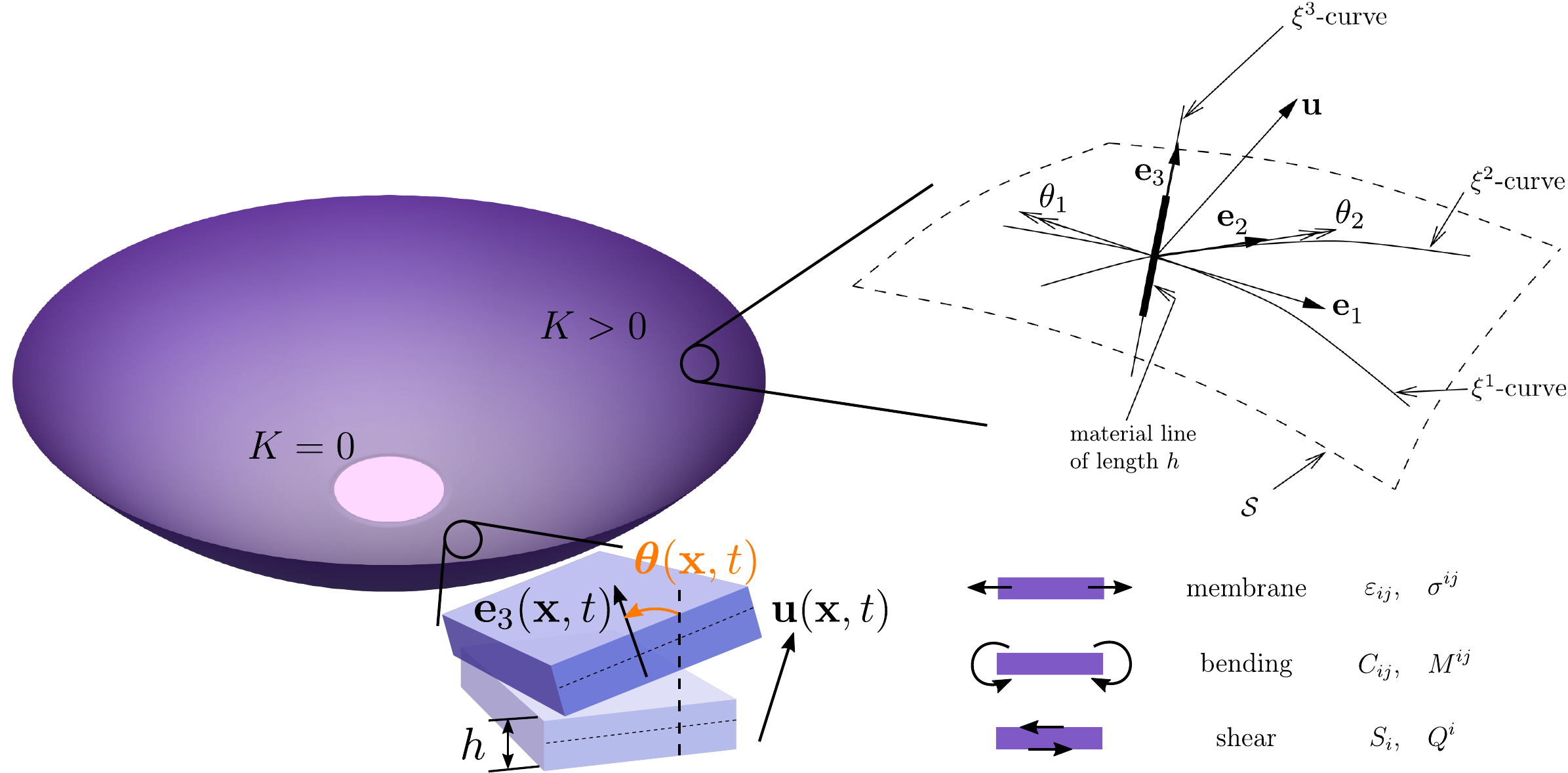}
    \caption{
        A simplified model of the steelpan as a thin elastic shell in the shape of a
        hemispherical bowl (positive Gaussian curvature \(K\)) with a flat central note
        (\(K=0\)). In the Naghdi shell theory, the deformations are described by the
        midsurface displacement \(\bb{u}\) and the rotation \(\vv{\theta}\) of material
        lines which are orthogonal to the midsurface in the undeformed configuration.
        The equations of motion (written in weak form in equation~\eqref{naghdi} and in
        strong form in equation~\eqref{naghdistrong}) take into account the elastic energy
        associated with membrane (stretching), bending and shear strains~\eqref{shellstrains}
        and stresses~\eqref{naghdistresses}
        (adapted from Shankar et al.~\cite{shankar2022}, and Chapelle and
        Bathe~\cite{bathechapelle}).
    }
\label{fig:geometry}
\end{figure}

We let \(\bb{U}(\vv{\xi})\) be the displacement of the material point of the shell which is
initially at \(\bb{R}_0(\vv{\xi})\) (the deformed position is then
\(\bb{R}=\bb{R}_0+\bb{U}\)).
To reduce the dimensionality of the model, we make the Reissner--Mindlin kinematic
assumption,
which specifies the displacement
\(\bb{U}\) of an arbitrary point of the shell in terms of two vector fields defined on the
midsurface: A displacement vector field \(\bb{u}\) and a rotation \((\theta_1, \theta_2)\)
of the unit normal. Specifically, we assume that the displacement takes the form
\begin{equation}
    \label{RMassumption}
    \bb{U} = u^{\alpha} \bb{e}_{\alpha} + \xi^3\theta^i\bb{e}_i.
\end{equation}
In words, the assumption states that a material line initially normal to the midsurface
remains straight and unstretched in the deformed state, but may be translated and rotated.
We see that \(\bb{u}\) is the translation of the midsurface while \(\theta_i\) are
the rotations of the material line around the axes defined by the tangent
vectors $\bb{e}_i$ (see figure~\ref{fig:geometry}).

The full nonlinear strain tensor is defined by
$E_{\alpha\beta}=(1/2)(\bar{g}_{\alpha\beta}-g_{\alpha\beta})$, where
$\bar{g}_{\alpha\beta}=\partial_\alpha\bb{R}\cdot\partial_\beta\bb{R}$ is the metric
of the deformed configuration. To leading order in variations along the thickness
(i.e. $\xi^3$), we obtain
\begin{equation}
    \label{straintensor}
    \begin{aligned}
        E_{ij} &= \varepsilon_{ij}+\xi^3C_{ij} + \mathcal{O}\big((\xi^3)^2\big),\\
        E_{i3} &= S_{i}, \\
        E_{33} &= 0,
    \end{aligned}
\end{equation}
where $\varepsilon_{ij}$ is the in-plane strain, $C_{ij}$ is the curvature or bending strain
and $S_i$ is the transverse shear strain, given by
\begin{subequations}
    \label{shellstrains}
    \begin{align}
        \varepsilon_{ij}&=\dfrac{1}{2}\left(\del_iu_j+\del_ju_i\right)-b_{ij} u_3,
        \label{membranestrain}\\
        C_{ij}&= \dfrac{1}{2}\left(\del_i\theta_j+\del_j\theta_i-b_i^k\del_ju_{k}-b_j^k\del_iu_{k}\right)
        + b_i^kb_{kj} u_3,\label{bendingstrain}\\
        S_i&=\dfrac{1}{2}\left(\theta_i+\partial_i u_3 + b_i^k u_k\right),\label{shearstrain}
    \end{align}
\end{subequations}
as depicted in figure~\ref{fig:geometry}\textit{b}.

\emph{Remarks.} (i) In the case of a flat plate, the second fundamental form \(b_{ij}\)
vanishes and equations~\eqref{straintensor}--\eqref{shellstrains} reduce to the usual strain tensor
of the Reissner--Mindlin plate model.
Note that in this case, the in-plane membrane strain tensor \(\varepsilon_{ij}\) depends only
on the in-plane displacement \(u_i\) and is uncoupled from the bending and shearing strains.
(ii) A displacement field of the form given in equation~\eqref{RMassumption} where
\(\theta_i=-\del_i u_3-b_i^k u_k\) is said to satisfy the
Kirchhoff--Love kinematic assumption.
It is stronger than the Reissner--Mindlin hypothesis as it
imposes that a material line which is initially normal to the midsurface remains normal in
the deformed state as well. From equation~\eqref{shearstrain}, this is equivalent to the
transverse shear strain $S_i$ vanishing. (iii) If the displacement field satisfies the
Kirchhoff--Love assumption and the shell is flat ($a_{ij}=\delta_{ij},\,b_{ij}=0$), we
obtain the Kirchhoff--Love plate model.

\subsection{Governing equations}
\label{subsec:eom}
We derive the dynamical Naghdi equations starting from the continuum mechanics of the
three-dimensional shell body. The constitutive relation for the 3D stress tensor
$\Sigma^{\alpha\beta}=H^{\alpha\beta\mu\nu}E_{\mu\nu}$ involves the elastic tensor (for an
isotropic material)
\begin{equation}
    \label{stiffnesstensor}
    H^{\alpha\beta\mu\nu}=\dfrac{Y}{2(1+\nu)}
    \left[g^{\alpha\mu}g^{\beta\nu} + g^{\alpha\nu}g^{\beta\mu} +
    \dfrac{2\nu}{(1-2\nu)}g^{\alpha\beta}g^{\mu\nu}\right],
\end{equation}
where $Y$ and $\nu$ are the Young's modulus and Poisson's ratio of the material,
respectively. Using the expression for the 3D metric tensor, equation~\eqref{3Dmetric},
along with the standard additional assumption that the normal stress $\Sigma^{33}$ vanishes
everywhere, we get
\begin{equation}
    \label{shellstress}
    \Sigma^{ij}=\tilde{\mathcal{A}}^{ijk\ell}E_{k\ell}, \quad
    \Sigma^{i3}=\dfrac{1}{2}\tilde{\mathcal{B}}^{ij}E_{j3},
\end{equation}
with the reduced elastic tensors
\begin{equation}
    \label{CD3d}
    \tilde{\mathcal{A}}^{ijk\ell}=\dfrac{Y}{2(1+\nu)}\left(g^{ik}g^{j\ell} +
    g^{i\ell}g^{kj}+\dfrac{2\nu}{(1-\nu)}g^{ij}g^{k\ell}\right), \quad
    \tilde{\mathcal{B}}^{ij}=\dfrac{2Y}{1+\nu}g^{ij}.
\end{equation}
Newton's second law for the 3D shell reads
\begin{equation}
    \label{strongeom}
    \rho\partial_t^2\bb{U}=\vv{\del}\cdot\vv{\Sigma}+\bb{F},
\end{equation}
where $\rho$ is the mass density and $\bb{F} = F^{\alpha} \bb{e}_{\alpha}$ is an external
force density on the shell. To obtain a weak form of the equation, suitable for the finite
element method, we introduce a test vector field $\bb{V}$ which satisfies the same kinematic
assumptions as $\bb{U}$, i.e., $\bb{V} = v^{\alpha} \bb{e}_{\alpha} + \xi^3\eta^i\bb{e}_i$,
and vanishes on the lateral boundary (corresponding to \((\xi^1, \xi^2) \in \partial
\Omega_0\)) where the displacement boundary condition \(\bb{U} = \bb{0}\) is prescribed.

Taking the inner product of equation~\eqref{strongeom} with $\bb{V}$ and integrating by parts,
we obtain
\begin{equation}
    \label{weakeom}
    \int_{\Omega}\dd V\rho\partial_t^2\bb{U}\cdot\bb{V}+\int_{\Omega}\dd
    V\;\vv{\Sigma}(\bb{U}):\bb{E}(\bb{V})=\int_{\Omega}\dd
    V\;\bb{F}\cdot\bb{V}+\int_{\partial\Omega}\dd
    S\;\left(\vv{\Sigma}(\bb{U})\cdot\hat{\vv{\nu}}\right)\cdot\bb{V}.
\end{equation}
where $\dd V=\sqrt{\det(\bb{g})}\dd\xi^1\dd\xi^2\dd\xi^3$ is the volume element and
$\hat{\vv{\nu}}$ is the unit outward normal along the boundary of the 3D shell, with $\dd
S$ the boundary area element.
The last term of equation~\eqref{weakeom} represents the effect of boundary tractions,
and can be divided into integrals over the upper and lower faces of the shell (corresponding
to $\xi^3 = \pm h/2$), and an integral over the lateral faces.
The upper and lower faces of the shell are free, meaning that
the traction $\vv{\Sigma}(\bb{U})\cdot\hat{\vv{\nu}}$ vanishes.
Moreover, \(\bb{V}=0\) on the lateral faces as previously noted.
Therefore, the last term of equation~\eqref{weakeom} vanishes for this choice of boundary
conditions.

Upon expanding the second term on the left-hand side of equation~\eqref{weakeom}, and
integrating over the thickness of the shell (to leading order in $\xi^3$, conversely $h$),
we obtain the dynamic Naghdi equations in weak form:
\begin{equation}
    \label{naghdi}
    \begin{gathered}
        \int_{\Omega_0}\dd A\;\rho h \partial_t^2 u_{\alpha} v^{\alpha} +
        \int_{\Omega_0}\dd A\;h\mathcal{A}^{ijk\ell}\left[\varepsilon_{ij}(\bb{u})\varepsilon_{k\ell}(\bb{v}) +
        \dfrac{h^2}{12}C_{ij}(\bb{u},\vv{\theta})C_{k\ell}(\bb{v},\vv{\eta})\right]\\
        + \int_{\Omega_0}\dd A\; h \mathcal{B}^{ij}S_{i}(\bb{u},\vv{\theta})S_j(\bb{v},\vv{\eta})
        = \int_{\Omega_0}\dd A\; h F_{\alpha} v^{\alpha},
    \end{gathered}
\end{equation}
where \(\dd A=\sqrt{\det(\bb{a})}\dd\xi^1\dd\xi^2\) is the area element on the midsurface and
\(\mathcal{A},\mathcal{B}\) are the restrictions of the elastic tensors \(\tilde{\mathcal{A}},
\tilde{\mathcal{B}}\)
onto the midsurface, i.e.,
\begin{equation}
    \label{CD2d}
    \mathcal{A}^{ijk\ell}=\dfrac{Y}{2(1+\nu)}\left(a^{ik}a^{j\ell}+a^{i\ell}a^{jk}+\dfrac{2\nu}{(1-\nu)}a^{ij}a^{k\ell}\right), \quad
    \mathcal{B}^{ij}=\dfrac{2Y}{(1+\nu)} a^{ij}.
\end{equation}
It remains to specify an appropriate space of functions \(\mathcal{V}\) in which to seek the
solution. Since the weak form of Naghdi's equations, equation~\eqref{naghdi}, involves only
first order derivatives, this will be a subspace of the Sobolev space
\((H^1(\Omega_0))^5\). In order to impose clamped (homogeneous Dirichlet) boundary
conditions, we take \(\mathcal{V} \defeq (H^1_0(\Omega_0))^5\), the subspace of functions
which vanish at the boundary.

For convenience, we will use the shorthand $u = (\bb{u}, \vv{\theta})$, $v = (\bb{v},
\vv{\eta})$, and let $k_m$, $k_b$ and $k_s$ denote the membrane, bending and shear terms
of equation~\eqref{naghdi}:
\begin{equation}
    \begin{aligned}
        k_m(u, v) &=
        \int_{\Omega_0}\dd A\;h\mathcal{A}^{ijk\ell}
        \varepsilon_{ij}(\bb{u})\varepsilon_{k\ell}(\bb{v})\\
        k_b(u, v) &=
        \int_{\Omega_0}\dd A\; \mathcal{A}^{ijk\ell}
        \dfrac{h^3}{12}C_{ij}(\bb{u},\vv{\theta})C_{k\ell}(\bb{v},\vv{\eta})\\
        k_s(u, v) &=
        \int_{\Omega_0}\dd A\; h \mathcal{B}^{ij}S_{i}(\bb{u},\vv{\theta})S_j(\bb{v},\vv{\eta}).
    \end{aligned}
\end{equation}
Lastly, we define the bilinear forms
\begin{equation}
    \begin{aligned}
        m(u, v) &= \int_{\Omega_0}\dd A\;\rho h u_{\alpha} v^{\alpha},\\
        k(u, v) &= k_m(u, v) + k_b(u, v) + k_s(u, v),
    \end{aligned}
\end{equation}
which we call the mass and stiffness form, respectively. With this notation in place,
the eigenvalue problem associated with~\eqref{naghdi} can be stated succinctly as:
Find pairs \((\omega^2, u)\) with \(\omega^2 > 0\),
\(u = (\bb{u}, \vv{\theta}) \in \mathcal{V} \defeq (H^1_0(\Omega_0))^5\) so that
\begin{equation}
    \label{naghdieigen}
    k(u, v) = \omega^2 m(u, v) \qquad \text{for all } v \in \mathcal{V}.
\end{equation}

For completeness, we mention that the Naghdi equations can also be written as a boundary
value problem in strong form as follows: Define the in-plane stress \(\vv{\sigma}\), bending
moment \(\vv{M}\) and transverse shear stress \(\bb{Q}\) by
\begin{equation}
    \label{naghdistresses}
    \begin{gathered}
        \sigma^{ij} = h \mathcal{A}^{i j k \ell} \varepsilon_{k \ell}\\
        M^{ij} = \frac{h^3}{12} \mathcal{A}^{i j k \ell} C_{k \ell}\\
        Q^{i} = \frac{h}{2} \mathcal{B}^{i j} S_j.
    \end{gathered}
\end{equation}
Then the weak form~\eqref{naghdi} is formally equivalent to the following system of
equations:
\begin{equation}
    \begin{alignedat}{2}
        \label{naghdistrong}
        \rho h \partial_t^2 u^{i} &= \nabla_j \sigma^{i j}
        - b^i_j Q^{j} - \nabla_j \big( b^{i}_{k} M^{k j} \big) + h F^{i},
        \qquad &&i = 1,2,\\
        \rho h \partial_t^2 u^{3} &= \nabla_i Q^{i} + b_{i j} \sigma^{i j}
        - b^{k}_{i} b_{k j} M^{i j} + h F^{3},\\
        0 &= Q^{i} - \nabla_j M^{i j}, && i = 1,2.
    \end{alignedat}
\end{equation}

\emph{Remarks.} (i) We note that in equation~\eqref{naghdi}, the bending term
$k_b$ is of higher order in the thickness $h$ compared to the stretching and shearing
terms, a general feature of shell models~\cite{bathechapelle}.
In the limit of small thickness,
modes with vanishing membrane and shear strains are energetically favorable. In the linear
theory, these pure-bending modes are characterized by $\vv{\varepsilon}=\bb{0}$, and
$\vv{\theta}=-(\vv{\del}u_3 + \bb{b}\cdot\bb{u})$. Existence of pure-bending displacements
depends on the boundary conditions as well as the midsurface reference geometry, in
particular the sign of the Gaussian curvature
$K = \det(\bb{b})/\det(\bb{a})$~\cite{bathechapelle}.
Isometric bending deformations are well understood for surfaces where $K$ has a constant
sign, but less is known for surfaces of mixed type where $K$ changes sign.
(ii) If we replace the Reissner--Mindlin kinematic assumptions with the stronger
Kirchhoff--Love assumption, we obtain instead the weak form of Koiter's
equations~\cite{bathechapelle}.
(iii) In the absence of curvature (i.e.\ \(b_{ij} = 0\)), the variational problem of
equation~\eqref{naghdi} splits into two decoupled equations: a
membrane problem for the in-plane displacement $(u_1, u_2)$ and the remaining variables
($u_3$, $\vv{\theta}$) satisfy the Reissner--Mindlin plate equations.
If we instead make the Kirchhoff--Love kinematic assumption, which for a flat plate states
that \(\theta_i = -\nabla_i u_3\), then the normal displacement \(u_3\) satisfies a simple
bending equation. If we choose orthonormal coordinates so that \(a_{ij} = \delta_{ij}\),
the equation is
\begin{equation}
    \rho h\partial_t^2 u^3 + B \del^4 u^3 = h F^{3}
\end{equation}
where \(B = Yh^3/[12(1-\nu^2)]\).
This is most easily
seen from the strong form equations~\eqref{naghdistrong}.

\section{Steelpan model and numerical simulations}
\label{sec:methodology}

\subsection{Numerical methods}
\label{subsec:num}
We discretize the Naghdi eigenvalue problem, equation~\eqref{naghdieigen}, in space using the
finite element method. As previously mentioned, the weak form of the Naghdi equations admits
solutions in $H^1$. However, it is well known that standard $H^1$-conforming finite element
methods suffer from numerical locking when applied to shell models, which leads to overstiff
behavior unless a very fine mesh is used. Several approaches to alleviate locking have been
proposed, see \cite{arnold1997,hale2018}
and references therein.
While no method has been rigorously shown to be locking free for all problems, numerical tests
suggest that many of the known methods can successfully treat locking in the Naghdi model.
Following \cite{hale2018}, we use a high-order partial selective reduced integration
(PSRI) method, a variation of the technique introduced in \cite{arnold1997}. This method was
shown in \cite{arnold1997} to converge uniformly with respect to the shell thickness under
some restrictive assumptions on the coefficients in the Naghdi model.
In the PSRI approach, second-order Lagrange
finite elements augmented by cubic bubble functions are used for the displacements
$(u_{\alpha})$, and second-order Lagrange finite elements are used for the rotations
$(\theta_i)$. The stiffness form $k$ in equation~\eqref{naghdieigen} is modified by splitting
the membrane term \(k_m\) and shear term \(k_s\) into a weighted sum of two contributions,
one of which is computed with a reduced integration. That is, we write \(k_m\) as
\begin{equation*}
    \alpha \int_{\Omega_0}\dd A\;h\mathcal{A}^{ijk\ell} \varepsilon_{ij}(\bb{u})\varepsilon_{k\ell}(\bb{v}) +
    (1-\alpha) \int_{\Omega_0}\dd A\;h\mathcal{A}^{ijk\ell}
    \varepsilon_{ij}(\bb{u})\varepsilon_{k\ell}(\bb{v})
\end{equation*}
and compute the second integral using a reduced quadrature rule of order 2.
The shear term \(k_s\) is similarly modified. The splitting parameter $\alpha$ is chosen
as $h^2/\delta^2 \approx 0.05$ where $\delta$ is a typical element circumradius for the
mesh, as suggested in~\cite{hale2018}.
Discretizing equation~\eqref{naghdieigen} in the manner just described yields a (generalized)
matrix eigenvalue problem which we solve using the SLEPc implementation of the Krylov-Schur
algorithm~\cite{slepc}.
The numerical method just described is implemented in a custom code based on the
FEniCS-Shells library\cite{fenicsshells} and using the open-source finite element computing
platform FEniCS\cite{fenics}.
The code is included in the supplementary material~\cite{steelpanSI}.

\subsection{Steelpan model and numerical experiments}
In a real steelpan, the crafting process introduces inhomogeneities in the thickness and
material properties. While these effects may influence the sound of the instrument, as we
show, they are not necessary for studying mode localization and can be neglected for our
purposes. For the same reason, the skirt of the pan is not included in our model.

To investigate mode confinement in a simplified model of the steelpan, we performed two
numerical experiments. In the first part of our analysis, we consider how the strength of
localization near a note varies with the pan geometry. In the second part, we test the
predictions of the landscape theory developed in \S\ref{sec:localization}. In each case, the
modes of the structure are computed by numerically solving the eigenvalue
problem~\eqref{naghdieigen} using the finite element method as described
in~\S\ref{subsec:num}. Since the localized modes under considerations are insensitive to
boundary conditions, we restrict the analysis to shells with a fully clamped boundary. In
all cases, we use a triangular mesh, with the mesh size chosen so that the modes of interest
have adequately converged; the number of elements is $\sim 15000$. We take the constant
thickness of the shell to be \(h = \SI{1}{\milli\meter}\), which is commonly used for these
instruments. In our simulations, we fix the radius of the pan at \(R_{\text{pan}} =
\SI{0.3}{\meter}\), approximately the radius of a traditional steelpan. The material
properties are \(Y = \SI{200}{\giga\pascal}\), \(\sigma = 0.3\) and \(\rho =
\SI{7850}{\kilo\gram\per\meter\cubed}\), typical for mild steel.

\begin{figure}
    \centering
    \includegraphics[width=0.95\textwidth, trim={0.00cm 0.00cm 0.00cm 0.00cm}, clip]{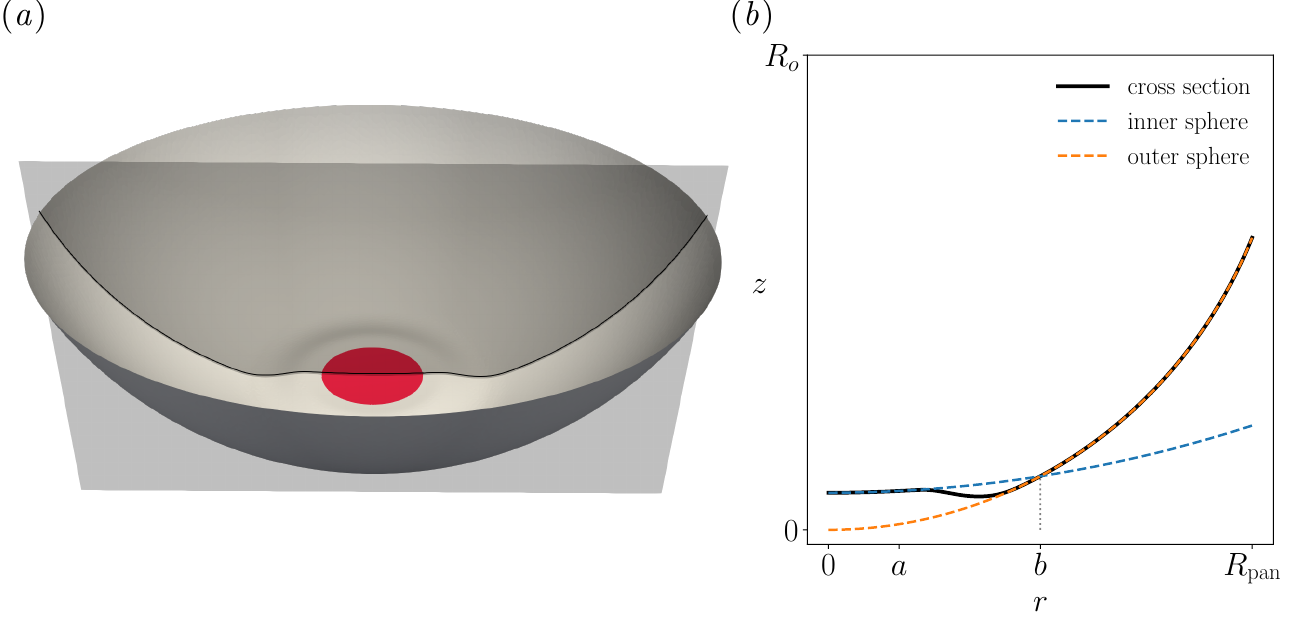}
    \caption{
        (\textit{a}) The idealized steelpan geometry with the diameter of the note indicated.
        (\textit{b}) Cross section of an idealized steelpan geometry which smoothly interpolates
        between an inner sphere of radius \(R_i\) for \(r \leq a\) and an outer sphere of
        radius \(R_o\) for \(r \geq b\). The width \(b-a\) of the transition
        region is exaggerated for clarity.
    \label{panshape}}
\end{figure}

In the first experiment, we consider how the localization strength varies with the
note and bowl curvatures.
For simplicity, we consider a radially symmetric geometry with a single note in the center,
shown in figure~\ref{panshape}\textit{a}.
The shape is chosen so that the inner note region and outer bowl regions each have constant
Gaussian curvature, which we define as \(\kappa_i^2\) and \(\kappa_o^2\) respectively.
Specifically, the inner note region \(0 \leq r < a\) and outer bowl region \(b <
r \leq 1\) are segments of spheres with radii \(R_i = \kappa_i^{-1}\)
and \(R_o = \kappa_o^{-1}\) respectively (figure~\ref{panshape}\textit{b}).
In the transition region \(a < r < b\) we choose a smooth interpolation between the inner
and outer spherical segments to ensure that the curvature is everywhere defined and
continuous. In our experiments, we take \(a = \SI{0.05}{\meter}\), \(b =
\SI{0.06}{\meter}\), so that the note area is less than 5\% of the total area of the pan.
As we vary the curvatures \(\kappa_i\) and \(\kappa_o\), we use two different measures to
quantify the localization strength of the so-called \((0,0)\) mode, which has no nodal
lines (see figure~\ref{fig:localization}\textit{b}).
The first is the inverse participation ratio (IPR), which for a mode \(u = (\bb{u},
\bb{\theta})\) of the Naghdi shell is defined by
\begin{equation}
    \label{iprdef}
    \operatorname{IPR} \defeq
    \operatorname{area}(\mathcal{S}) \int \norm{\bb{u}}^4 \dif S
    \left(\int \norm{\bb{u}}^2 \dif S\right)^{-2}.
\end{equation}
If \(\norm{\bb{u}}\) is approximately constant on a subregion \(\mathcal{D} \subset
\mathcal{S}\) and vanishing outside of \(\mathcal{D}\), then the participation ratio is
\(\operatorname{area}(\mathcal{S})/\operatorname{area}(\mathcal{D})\).
We also use another, somewhat ad hoc measure of localization, which measures the proportion
of the mode that lives in the note region \(\mathcal{N}\) in an \(L^1\)-norm sense.
If we assume that the mode has been normalized so that \(\norm{\bb{u}}\) has
\(L^{1}\)-norm 1, then this is simply
\begin{equation}
    \label{lnrdef}
    \norm{\bb{u}}_{L^1(\mathcal{N})} \defeq \int_{\mathcal{N}} \norm{\bb{u}} \dif A.
\end{equation}
We refer to this measure as the \(L^1\)-norm ratio (LNR). We note that while the LNR
is bounded above by 1 (achieved by any mode that vanishes outside \(\mathcal{N}\)),
the IPR can be arbitrarily large.
These two measures of localization strength are complimentary:
The LNR is more sensitive to small amplitude oscillations outside of the note region, but
unlike the IPR, it gives no information about how narrowly peaked the mode is inside of
\(\mathcal{N}\).

In the second experiment, we use the vector landscape of \S\ref{sec:localization}
to study mode confinement in two idealized steelpan geometries, each consisting of a
hemispherical bowl with one or more flat note regions where the curvature vanishes.
The (nondimensionalized) Gaussian curvature of the bowl is \((h \kappa_o)^2 = \num{8.2e-4}\)
in both cases.
The first pan shape, depicted in the top left of figure~\ref{fig:landscape}\textit{a}, has a
single, circular note of radius \(\SI{0.05}{\meter}\) in the center. The second geometry,
which more closely resembles the elaborate designs of real steelpan instruments, has four
notes. Three of those are elliptical inner notes of varying size and eccentricity, placed
close together near the center. The fourth note is an approximately rectangular outer note
near the boundary (figure~\ref{fig:landscape}\textit{b}).
For each of the two pan shapes, we compute the vector landscape \(\vv{\mathcal{L}}\) and
several low-frequency eigenmodes. To compute the value of \(\vv{\mathcal{L}}\) at a point
\(y\) on the shell, we solve the Dirichlet problem \(L \bb{G}(\cdot, y) = \delta_y I\) for
\(\bb{G}(\cdot, y)\) using the finite element method and compute the integral in
equation~\eqref{vectorlandsape}.

\section{Results}
\label{sec:results}

\subsection{Effect of geometry on mode localization}
\label{results:localization}
\begin{figure}
    \centering
    \includegraphics[width=0.9\textwidth, trim={0cm 0.00cm 0.00cm 0.00cm}, clip]{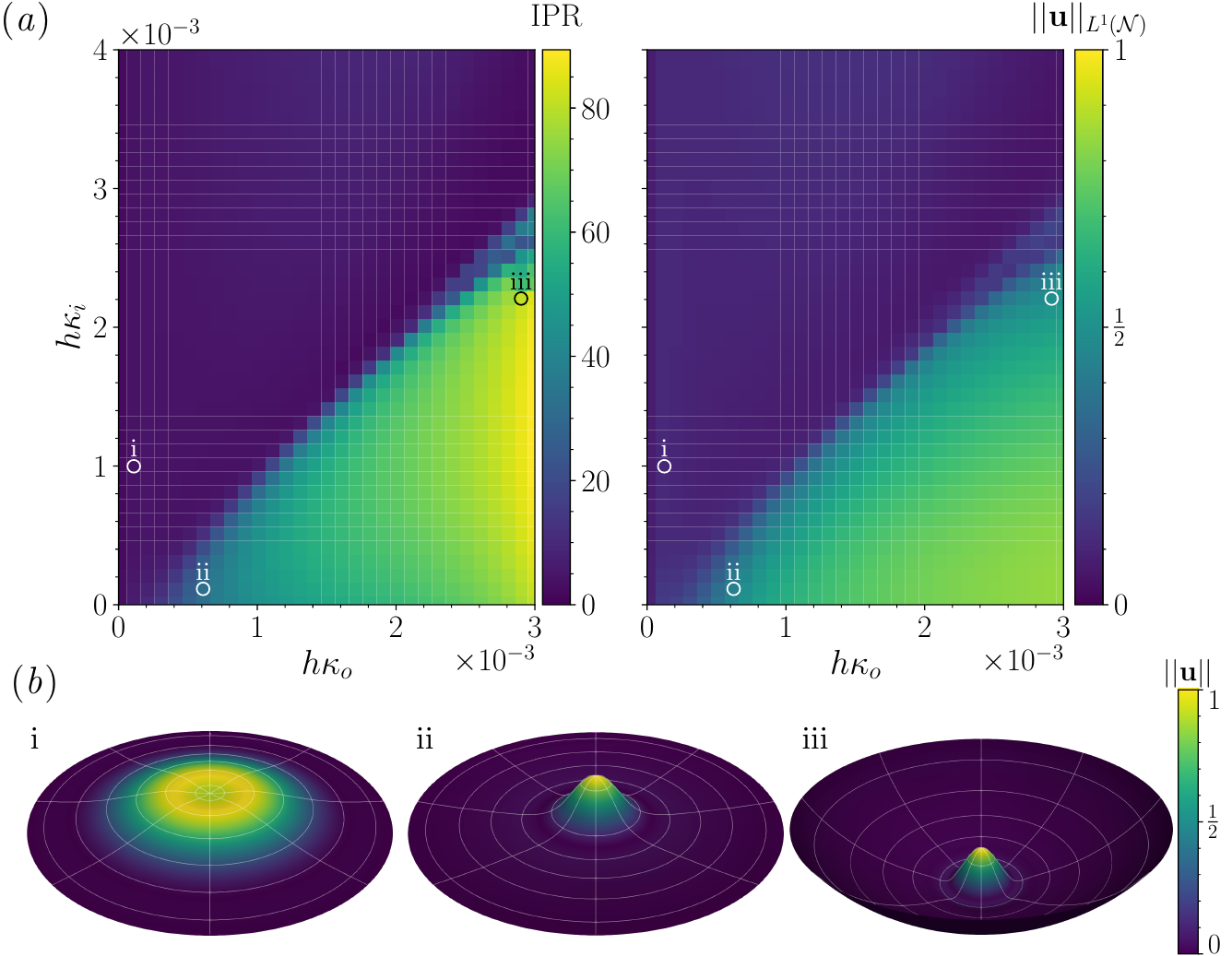}
    \caption{
        (\textit{a}) Inverse participation ratio (left) and \(L^1\)-norm ratio (right) of the
        \((0,0)\) mode as a function of the nondimensionalized note and bowl curvatures.
        A sharp transition from localized to extended modes is seen near the diagonal
        \(\kappa_o = \kappa_i\).
        (\textit{b}) Representative mode shapes corresponding to the points marked in (a).
        Color indicates the normalized displacement and grid lines are shown for clarity.
        While modes ii and iii haves approximately the same \(L^1\)-norm ratio, the latter
        has a much larger inverse participation ratio.
    \label{fig:localization}}
\end{figure}

Figure~\ref{fig:localization}\textit{a} shows the IPR and LNR of the \((0,0)\)
mode as a function of the bowl and note curvatures.
In the region \(\kappa_i > \kappa_o\), both localization measures are small and
approximately constant, indicating that the mode is extended over most of the pan.
Near the diagonal \(\kappa_i = \kappa_o\), there is a sharp transition from extended to
localized modes which is reflected in both the IPR and LNR.
Three representative mode shapes, corresponding to the indicated points i-iii in parameter
space, are shown in figure~\ref{fig:localization}\textit{b}.

In the localized region, \(\kappa_i < \kappa_o\), the two localization measures diverge.
The \(L^1\)-norm ratio seems to increase monotonically with the difference \(\kappa_o - \kappa_i\)
between the outer and inner curvatures. The IPR, in contrast, is approximately independent
of \(\kappa_i\) in this region, but increases with \(\kappa_o\).
This difference can be seen by comparing modes ii and iii
in figure~\ref{fig:localization}\textit{b}, which are both confined to the note region
with similar \(L^1\) norm ratios, but have substantially different values of the IPR.
Therefore, the inverse participation ratio gives more fine-grained information about the
shape of the localized mode in the note region.

\subsection{Illustration of the vector landscape}
\label{results:landscape}
\begin{figure}
    \centering
    \includegraphics[width=0.8\textwidth, trim={0.08cm 0.06cm 0.05cm 0.05cm}, clip]{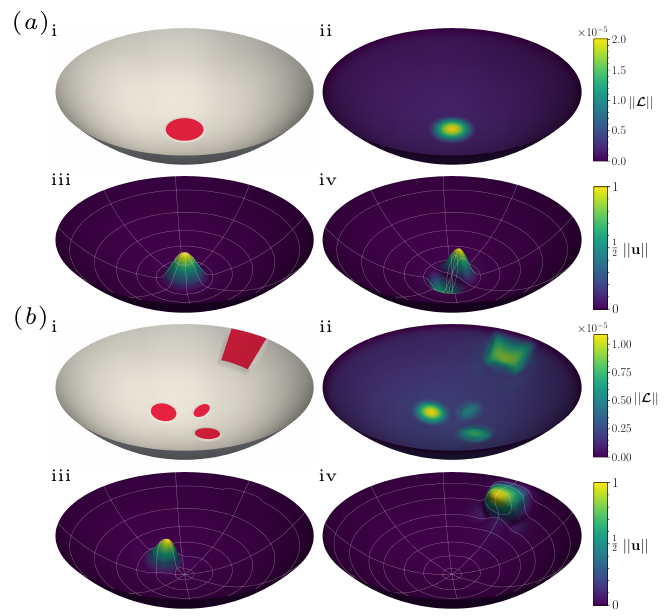}
    \caption{
        The vector landscape and several eigenmodes were numerically computed for two
        steelpan-like geometries, shown in figures (\textit{a},\textit{b}) i with
        the flat note regions highlighted in red.
        Figures (\textit{a},\textit{b}) ii  show the norm of the vector landscape
        \(\vv{\mathcal{L}}\) of equation~\eqref{vectorlandsape} for the two geometries. The first two numerically computed
        eigenmodes are shown in iii-iv, exhibiting strong localization to the regions
        predicted by the vector landscape. Color indicates the normalized displacement
        (white grid lines are shown for clarity).
        The geometry in figures (a) has a single circular node in the center with
        \(a = 0.05\)\,m, \(b = 0.06\)\,m, \(R_i = 350\)\,m (see figure~\ref{panshape}).
        The geometry in (b) has four flat notes \(K=0\):
        A circular note of radius \(0.03\)\,m
        centered at \((x,y) = (-0.075, 0)\)\,m; an elliptical note with semi-major axis
        \(0.03\)\,m and eccentricity \(\sqrt{2}/2\) centered at
        \((x,y) = (0.075, 0)\)\,m;
        an elliptical note with semi-major axis \(0.23\)\,m and eccentricity
        \(\sqrt{2}/2\) centered at \((x,y) = (0, 0.075)\)\,m; a note in the shape of
        a polar rectangle defined by \(r \geq 0.215\)\,m, \(\abs{\theta} \leq
        \pi/18\) in polar coordinates.
        For both geometries, the bowl has radius of curvature \(R_o = 0.35\)\,m.
    \label{fig:landscape}}
\end{figure}

The vector landscape and several lowest eigenmodes were computed for the two shell geometries
described in \S\ref{sec:methodology}. The results for the single-note pan are shown in
figure~\ref{fig:landscape}\textit{a}. The norm \(\norm{\vv{\mathcal{L}}}\) of the vector
landscape is almost vanishing outside of the note region, where it has a sharp peak.
Based on this, we predict that the shell supports one or more localized modes in the note
region, and only high-frequency modes can have substantial amplitude in the outer bowl
region. This is confirmed in the lower half of figure~\ref{fig:landscape}\textit{a}, which
depicts the two lowest-frequency modes of the shell. The pan has several more
strongly localized modes which are not shown here.

Figure~\ref{fig:landscape}\textit{b} shows the more interesting case of a steelpan with
multiple notes of varying size and shape. The vector landscape \(\norm{\vv{\mathcal{L}}}\)
has several peaks, each located at note region where the curvature vanishes.
The tallest peak coincides with the large, circular inner note, followed by a peak at the
rectangular outer note. As explained in \S\ref{sec:localization}, the landscape
inequality~\eqref{vectorlandscapeineq} guarantees that any low-lying modes must be confined
to the union of the notes regions.
While this is an interesting result, it raises the question of whether the vector landscape
has the same predictive power as the scalar landscape, which is much stronger than what is
guaranteed by the inequality alone.
In fact, we observe that each mode is strongly localized to a \emph{single} note region and
that the mode shape is reflected in the shape of the corresponding peak of the landscape.
Moreover, the relative heights of the peaks is an indicator of the order in which the
localized modes appear. For example, the lowest-frequency mode is localized near the tallest
peak of the landscape.

\section{Conclusions}
The possibility of creating independent, geometrically tunable localized modes has been
utilized by makers of steelpans since the early 20th century, but a general quantitative
theory for how geometry can lead to effective localization has so far been missing. Inspired
by this spectral problem for the steelpan, we first generalized a recent geometric theory
for Anderson localization for linear scalar operators to a vector-valued version. We then
used the resulting localization landscape theory to predict the locations and order of
localized modes in doubly-curved elastic shells by solving a single Poisson-like problem
rather than an eigenvalue problem. Our results  show an interesting connection between
curvature and mode confinement in doubly curved elastic shells: shells with flat regions
separated by regions of positive Gaussian curvature support localized modes, and the
strength of localization increases with curvature. It might be pertinent to note that our
results are size and material independent, as the geometric theory of elastic shells is only
predicated  on having a large aspect ratio ($R/h \gg 1$). Thus they apply equally to micro-
or nanoscale systems, and and suggestive of applications to high quality
resonators~\cite{craighead2000}.  More generally, our generalization of the landscape theory
to vector-valued systems opens up applications to a much larger class of systems than
previously possible.

\begin{acknowledgments}
    We thank the NSF MRSEC 2011754, DMR 1922321, the Henri Seydoux Fund, and the Simons
    Foundation for partial financial support.
\end{acknowledgments}

\bibliography{references}

\end{document}